\documentclass[10pt,conference]{IEEEtran}
\usepackage[left=0.7in, right=0.73in, top=0.7in, bottom = 0.98in]{geometry}
\usepackage{graphicx}
\usepackage{setspace}
 \usepackage{comment}
\usepackage{adjustbox} % for adjusting 
\usepackage{enumitem}
\usepackage[noadjust]{cite}
\usepackage[linesnumbered,ruled]{algorithm2e}
\usepackage{amsmath}
\usepackage{lipsum}
\usepackage{mathtools}
\usepackage{cuted}
\usepackage{array}
\usepackage{subfigure}
\usepackage{amssymb}
\usepackage{caption}
\usepackage{multirow}
\usepackage{makecell}
\usepackage{algorithmic}
\usepackage{xcolor}
\def\BibTeX{{\rm B\kern-.05em{\sc i\kern-.025em b}\kern-.08em
ThedeltaT\kern-.1667em\lower.7ex\hbox{E}\kern-.125emX}}

\SetCommentSty{mycommfont}
\usepackage{array,multirow,makecell}
\SetKwInput{KwInput}{Input}    
\SetKwInput{KwRepeat}{Repeat} % Set the Input
\SetKwInput{KwOutput}{Output}    

\begin{document}

%\title{Gradient-based Meta Learning for Beyond Diagonal-RIS-aided uplink Rate-Splitting Multiple Access}

\title{Coordinated Multipoint Transmission in Pinching Antenna Systems}
\author{Ali Amhaz, Shreya Khisa, Mohamed Elhattab, Chadi Assi and Sanaa Sharafeddine \vspace{-0.6cm}}

\maketitle

\begin{abstract} 
We study a coordinated multi-point (CoMP) transmission where two base stations (BSs), each supported by a pinching antenna system (PASS), are deployed to jointly serve communication users under spatial division multiple access (SDMA) technology. Pinching Antenna technology was introduced as a promising solution to overcome the large-scale fading that has been shown to be an impediment in multiple-input multiple-output (MIMO) systems. To realize the advantages of this technology in CoMP systems, which suffer from an upperbound rate limitation when traditional uniform linear arrays (ULAs) are adopted, we formulate an optimization problem with the aim of maximizing the achievable sum rate by jointly determining the transmit beamforming vectors and pinching locations on the waveguides while respecting the quality of service (QoS) requirements of users. This problem is inherently non-convex due to the strong coupling among its decision parameters, making it challenging to solve using traditional optimization methods. Thus, we utilize a gradient-based meta-learning (GML) strategy specifically designed for large-scale optimization tasks. Finally, numerical analysis demonstrates the effectiveness of the proposed GML approach, achieving 92\% of the optimal solution, and the superiority of the solution presented compared to other benchmarks. In addition, it achieves a higher upper bound on the achievable rate compared to conventional CoMP systems.

\end{abstract}

\section{Introduction}
As we steadily advance towards the sixth-generation (6G) of wireless networks, it is expected that the communication landscape will undergo a massive transformation characterized by a rapid surge in connected devices, higher user density, and more demands for unprecedentedly higher data rates. This evolution will be further driven by the Internet of Everything (IoE) paradigm, which seeks to interconnect billions of users and smart devices, thereby requiring networks that can deliver ultra-high data rates, seamless global connectivity, and extremely low latency \cite{wang2023road}. A key enabling technology that has emerged for previous generations of wireless networks reaching up until 5G networks was multiple-input multiple-output (MIMO) systems due to their ability to improve the degrees of freedom (DoFs). The multiple antenna architecture on both the transmitter and receiver sides allows for an increase in user connectivity, spectral efficiency, and reliability \cite{he2021cell}. These significant advantages are achieved through the enhancement in spatial diversity and multiplexing gains, mitigating the interference effect at the users and improving the wireless coverage, as shown in several studies \cite{10851455}, \cite{8097026}.    

Although MIMO systems have established themselves as the cornerstone of modern wireless communications, their performance can still degrade due to inter-cell interference (ICI) generated by signals from neighboring cells, particularly in scenarios where users are located near cell edges. Coordinated multi-point (CoMP) technology was proposed to tackle these challenges. CoMP operates by allowing multiple base station (BSs) to cooperate coherently in serving users, guaranteeing a type of enhanced spatial diversity and improving the signal received at the users \cite{8097026}. Furthermore, CoMP exploits channels from multiple transmission points to enhance user multiplexing efficiency and effectively mitigate ICI. The authors in \cite{9737471} investigated a system model that employs a joint transmission (JT) CoMP architecture to serve cell-edge users, leveraging user cooperation to further enhance system performance. However, the spectral efficiency achievable at the user end is inherently limited by an upper bound, mainly due to persistent interference, even under high transmission power. This interference originates from signals intended for other users and cannot be fully mitigated, particularly in space division multiple access (SDMA) systems, where multiple users are simultaneously served over the same frequency resources through spatial separation.

Recently, pinching antennas (PAs) technology started to attract significant focus from the research community due to its ability to control the wireless transmission medium, similar to its preceding reflective intelligent surface and movable antenna technologies. First unveiled by NTT DOCOMO at the Mobile World Congress (MWC) 2021, PAs represent a novel reconfigurable antenna paradigm based on dielectric waveguide technology \cite{suzuki2022pinching}. In this architecture, electromagnetic signals propagate along low-loss dielectric waveguides, while the so-called “pinching points” allow signals to be selectively radiated into free space at desired positions along the waveguide. In \cite{10896748}, the authors exploited PAs with the goal of maximizing the achievable rate of the users by determining the optimal pinching locations and demonstrating advantages over conventional
fixed-location antenna benchmarks. Owing to its promising capabilities in controlling wireless channels, particularly in mitigating large-scale fading over long distances, we argue that this technology may offer an effective solution to the limitations of CoMP systems. In contrast, conventional MIMO systems often suffer from severe fading caused by non-line-of-sight (NLoS) links, high system complexity, substantial channel estimation overhead, and elevated implementation costs, all of which significantly degrade overall system performance.

To the best of our knowledge, no previous work examined the utilization of PAs technology in CoMP systems as a way to mitigate the limitations introduced above, and we argue that PAs may bring many benefits to the CoMP systems, which we intend to evaluate. Hence, we formulate an optimization problem with the objective of maximizing the achievable sum rate of the users while meeting the quality of service (QoS) requirements. This is done by optimizing the transmit beamforming vectors at the BSs involved in CoMP transmission, along with the pinching locations on the different waveguides. In addition, because of the non-convex nature of this optimization, we use a gradient-based meta-learning (GML) approach that can handle large-scale problems.  
%Owing to these promising properties and potentials in controlling the wireless channels, particularly the large-scale fading over long distances, it can provide a clever solution to the limitations of CoMP systems away from MIMO systems suffer from significant fading due to the non-line-of-sight (NLoS) links and the additional complexity, heavy channel estimation overhead, and  increased cost.  
%
%
\section{System Model}
\begin{figure} % the * makes it span across two columns
    \centering    \includegraphics[width=0.5\textwidth]{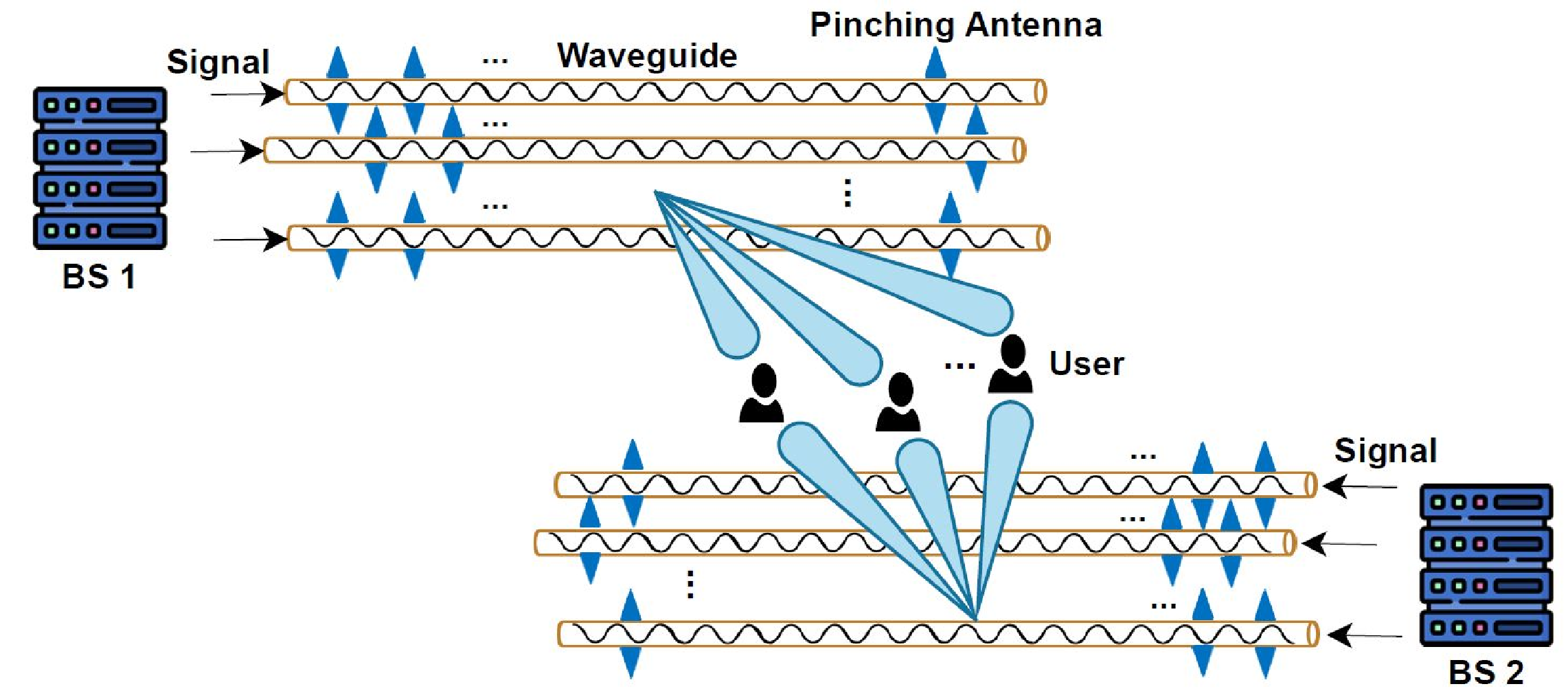}
    \caption{System Model}
    \label{fig:placeholder}
\end{figure}
In this work, a PA-enabled CoMP system comprises two BSs denoted as BS1 and BS2, jointly cooperating to serve a set of $K$ users, denoted as $\mathcal{K} = \{1, \dots, K\}$, as depicted in Fig. 1. BS1 and BS2 are equipped with $N_1$ and $N_2$ waveguides, where $N_1$ and $N_2$ denote the number of waveguides used by each BS, respectively. Each is connected to its corresponding BS using an RF chain. We define the set of waveguides attached to BS$b$, $\forall b\in \{1,2\}$ as $\mathcal{N}_b = \{1,2, \dots, n_b, \dots, N_b\}$. The waveguides of both BSs are assumed to be parallel to each other and aligned along the x-axis, where the height of waveguide $n_b$ at BS$b$ is denoted by $A_{n_b}$, with $b \in \{1, 2\}$. Along the y-axis, the coordinate of the $n_b$-th waveguide is denoted by $y_{n_b}$. Consequently, each waveguide extends from its feeding point $\boldsymbol{O}_{n_b} = [0, y_{n_b}, A_{n_b}]^T$ to $\boldsymbol{f}_{n_b} = [\mathcal{D}_{n_b}, y_{n_b}, A_{n_b}]^T$, where $\mathcal{D}_{n_b}$ represents the spanning distance of the $n_b$-th waveguide. Furthermore, the $n_b$-th waveguide is equipped with $P_{n_b}$ PAs, whose coordinates are given by $\boldsymbol{p}_{{n_b},p} = [x_{n_b, p}, y_{n_b}, A_{n_b}]^T$ for the $p$-th PA on the $n_b$-th waveguide. Finally, the location of the user $k$, $\forall k \in \mathcal{K}$ is denoted by $\boldsymbol{u}_k = [x_k, y_k, 0]^T$.

%\subsection{Downlink Channel}
By defining the complex information symbols intended for different users as $\textbf{c}_b=[c_1,\dots,c_K] \in \mathbb{C}^{K \times 1}$ and satisfying $\mathop{\mathbb{E}} \{\boldsymbol{c}\boldsymbol{c}^H\} = \boldsymbol{I}_K$, the overall signal at BS $b$ can be defined as $\boldsymbol{s_b}$=$\boldsymbol{W_b}\boldsymbol{c_b}$ and with a more detailed expression as: 
\begin{equation}
    \textbf{s}_b=\sum_{k \in \mathcal{K}}\textbf{w}_{k,b}{c}_{k,b}, \forall b \in \{1, 2 \},
\end{equation}
where $\boldsymbol{W}_b=[\boldsymbol{w}_{1,b},\dots,\boldsymbol{w}_{K,b}] \in \mathbb{C}^{N_b \times K}$ represents the transmit beamforming matrix and $\textbf{w}_{k,b} \in \mathbb{C}^{N_b \times 1}$ denotes the beamforming vector dedicated to the user $k$. Now, following the assumption that the in-waveguide propagation is lossless, the propagation channel in the $n_b$-th waveguide can be expressed as:
%
%\frac{P_{BS,b}}{N_b}
\begin{align}
&\textbf{g}_{n_b}=\sqrt{
\delta_{\text{eq}}
}\left[e^{\frac{-2\pi}{\lambda_g}\|\boldsymbol{p_{{n_b},1}}-\textbf{O}_{n_b}\|},\dots,e^{\frac{-2\pi}{\lambda_g}\|\boldsymbol{p_{{n_b},P_{n_b}}}-\textbf{O}_{n_b}\|}\right]^T \nonumber \\
&\in \mathbb{C}^{N \times 1},
\end{align}
where the guided wavelength, denoted by $\lambda_g = \frac{\lambda}{n_{\mathrm{eff}}}$, is determined by the effective refractive index $n_{\mathrm{eff}}$ of the dielectric waveguide, which characterizes the propagation of the electromagnetic wave within the medium and $\lambda$ representing the wavelength \cite{zhao2025waveguidedivisionmultipleaccess}. Additionally, $0 < \delta_{\text{eq}} \leq \frac{1}{P_{n_b}}  $ is the equal-power ratio where an equal proportion of the BSs' power is assigned to the different radiating antennas \cite{wang2025modelingbeamformingoptimizationpinchingantenna}. Next, the channel between the PAs of the waveguide $n_b$ and user $k$ can be defined as: 
\begin{equation}
\textbf{h}_{n_b,k}=\left[\frac{\eta e^{-j\frac{2\pi}{\lambda}\|\boldsymbol{p_{{n_b},1}}-\textbf{u}_k\|}}{\|\boldsymbol{p_{{n_b},1}}-\textbf{u}_k\|},\dots,\frac{\eta e^{-j\frac{2\pi}{\lambda_c}\|\boldsymbol{p_{{n_b},P_{n_b}}}-\textbf{u}_k\|}}{\|\boldsymbol{p_{{n_b},P_{n_b}}}-\textbf{u}_k\|}\right]^H,
\end{equation}
%
\begin{comment}
Here, $\eta \in \mathbb{R}$ is the channel gain accounting for he free space path loss factor.
\begin{equation}
    \textbf{h}_{t,k}=\boldsymbol{\beta}_t\hat{\textbf{h}}_{t,k}
\end{equation}
$[\boldsymbol{\beta}_t=\beta_{t,1}, \beta_{t,2}, \dots, \beta_{t,N}]^T$ antenna activation vector where $\beta_{t,n} \in \{0,1\}$.   
\end{comment}

\noindent where $\eta \in \mathbb{R}$ denotes the channel gain, which encapsulates the effects of free-space path loss as well as the radiation characteristics of the PAs. Now, let $\boldsymbol{s}_{n_b}$ represent the signal injected into the $n_b$-th waveguide, corresponding to the $n$-th component of $\boldsymbol{s}_b$. Hence, the signal which will be received at user $k$ can be denoted as:
\begin{equation}
y_k= \sum_{n_b=0}^{N_1} \textbf{h}_{n_b, k}^H\textbf{G}_1\textbf{s}_{n_b}+ \sum_{n_b=0}^{N_2} \textbf{h}_{n_b, k}^H\textbf{G}_2\textbf{s}_{n_b}+n_k,
\end{equation}
and can be expressed in a more detailed format as:
\begin{align}
&y_k=\textbf{H}_{ k,1}^H\textbf{G}_1\sum_{k \in K}\textbf{w}_{k,1} c_{k,1} \nonumber + \textbf{H}_{ k, 2}^H\textbf{G}_2\sum_{k \in K}\textbf{w}_{k,2} c_{k,2}  +n_k,  
\end{align}
where the first term corresponds to the transmission of BS1, and the second to that of BS2, and $n_k \in \mathcal{CN}(0, \sigma_k^2)$ denotes the additive white Gaussian noise (AWGN) with zero mean and variance $\sigma_k^2$. The terms $\textbf{H}_{k,b}$ and $\textbf{G}_b$ are defined as follows:
\begin{equation}
    \textbf{H}_{k,b}=[\textbf{h}_{1,k}^T,\dots,\textbf{h}_{N_b,k}^T]^T \in \mathbb{C}^{N_b P_{n_b} \times 1},
\end{equation}
\begin{equation}
    \textbf{G}_b=\textrm{diag}[\textbf{g}_1,\dots,\textbf{g}_{N_b}] \in \mathbb{C}^{N_b P_{n_b} \times N_b}.
\end{equation}
Thus, the achievable sum rate at user $k$ due to the coherent transmission from both BSs is expressed as follows:
\begin{align}
&R_{k}=\log_2\left(1+\frac{|\textbf{H}_{1, k}^H\textbf{G}_1\textbf{w}_{k,1} + \textbf{H}_{2,k}^H\textbf{G}_2\textbf{w}_{k,2}|^2}{I_1+I_2+\sigma_k^2}\right),
\end{align}
where the interference terms $I_1$ and $I_2$ can be expressed as follows:
\begin{equation}
   I_1 =  \sum_{k'\in \mathcal{K}, k' \neq k}|\textbf{H}_{1,k}^H\textbf{G}_1\textbf{w}_{k',1}|^2,
\end{equation}

\begin{equation}
   I_2 =  \sum_{k'\in \mathcal{K}, k' \neq k}|\textbf{H}_{2,k}^H\textbf{G}_2\textbf{w}_{k',2}|^2.
\end{equation}

\section{Problem Formulation}
With the objective of maximizing the sum rate, we formulate an optimization problem that jointly determines the transmit beamforming vectors $\boldsymbol{W}_1 = \{ \boldsymbol{w}_{k,1}, \forall k \in \mathcal{K} \}$ for BS1 and $\boldsymbol{W}_2 = \{ \boldsymbol{w}_{k,2}, \forall k \in \mathcal{K} \}$, along with the PAs positions $\boldsymbol{P}_1 = \{ \boldsymbol{p}_{n_1,p}, \forall 0 \leq  n_1 \leq N_1 \}$ and $\boldsymbol{P}_2 = \{ \boldsymbol{p}_{n_2,p}, \forall 0 \leq n_2 \leq N_2 \}$ corresponding to BS1 and BS2, respectively, while respecting the QoS requirements. In light of this, the problem can be formulated as follows:
\allowdisplaybreaks
\begin{subequations}
\label{prob:P1}
\begin{flalign}
 &\mathcal{P}_1: \max_{\substack{\textbf{P}_1,\textbf{P}_2, \,\textbf{W}_1, \textbf{W}_2}} \quad \sum_{k \in \mathcal{K}} R_{k}, \label{const1}\\
 &\text{s.t.} \quad \textrm{Tr} (\textbf{W}_b\textbf{W}_b^H) \le P_{BS,b}, \forall b \in \{1,2\}, \label{c0}\\
 &\qquad R_k \ge R_{th}, \forall k \in \mathcal{K}, \label{c1}\\
 &\qquad p_{n_b,p} - p_{n_b,p'}  \ge DS, \; p \neq p', \;  \forall 0 \leq p, p'  \leq P_{n_b}, \nonumber\\
 &\qquad\qquad \forall 0 \leq n_1 \leq N_1, \; \forall 0 \leq n_2 \leq N_2, \label{c2}\\
 &\qquad p_{n_b,p} \in \mathcal{D}_{n_b}. \label{c3}
\end{flalign}
\end{subequations}
where $P_{BS,b}$ is the power budget for BS $b$, $R_{th}$ is the rate threshold at the users, and $DS$ represent the minimum spacing between the PAs. The constraint \eqref{c0} is to respect the power budget at the different BSs. \eqref{c1} is for the QoS requirements for the different users, while  \eqref{c2} ensures a minimum distance between the PAs and \eqref{c3} specifies the operating range along the waveguide. This problem demonstrated non-convexity due to high coupling between the different variables, rendering a complex problem that is hard to tackle using traditional optimization solvers.
\section{Solution Approach}
\begin{figure} % the * makes it span across two columns
    \centering
    \includegraphics[width=0.45\textwidth]{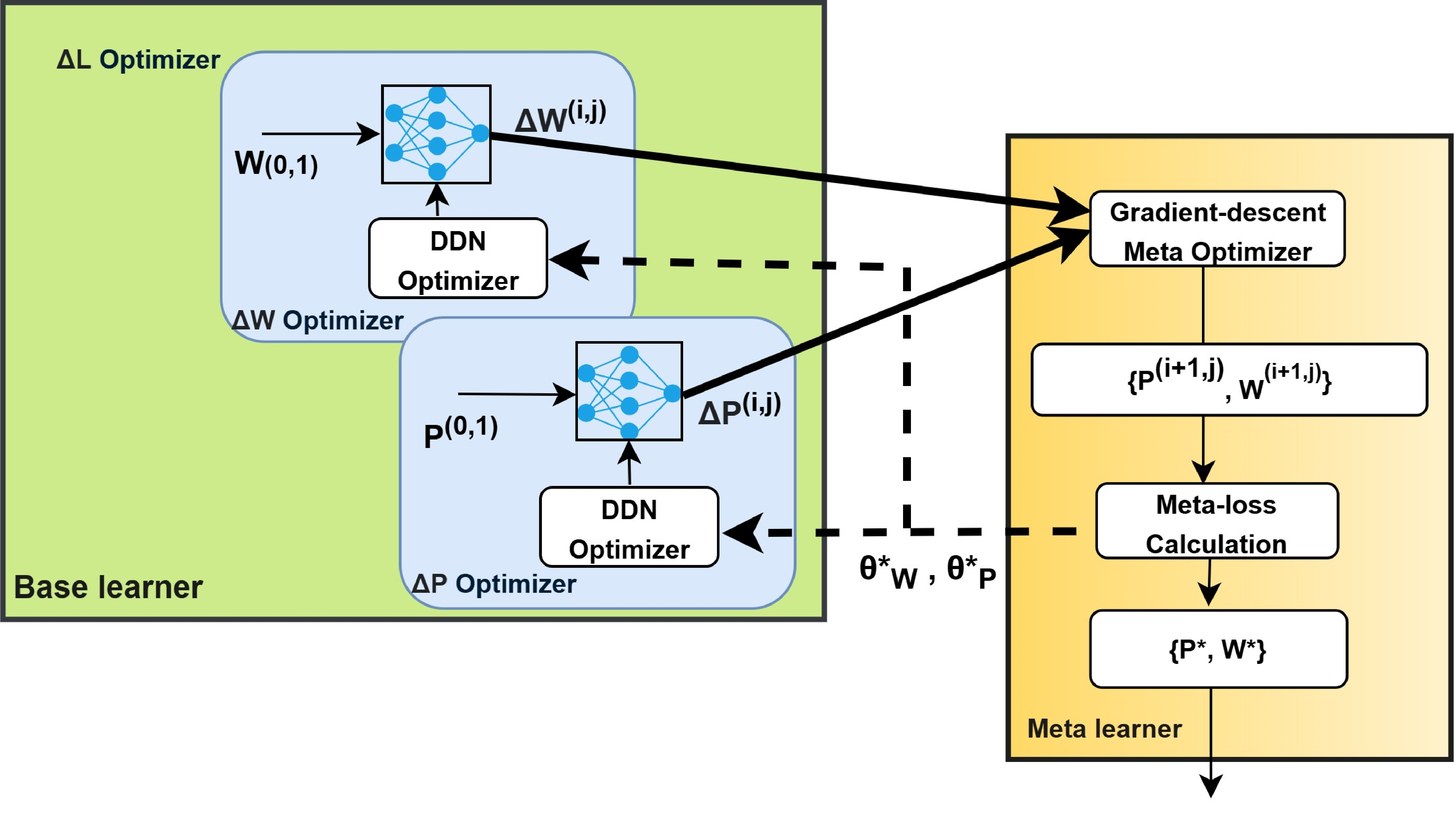}
    \caption{GML model architecture.}
    \label{GML_Model}
\end{figure}
 We rely on a GML algorithm, developed to handle large-scale and complex optimization problems. GML operates by obtaining the gradients of the different variables as $\Delta_{\boldsymbol{W}}R_{\boldsymbol{W}}$ and $\Delta_{\boldsymbol{P}}R_{\boldsymbol{P}}$, based on the objective function \eqref{const1} referred to as R, and feeding it to two neural networks which are responsible for generating gradients $\Delta\boldsymbol{W}$ and  $\Delta \boldsymbol{P}$ that will add over the initial values of $\boldsymbol{W} = \{\boldsymbol{W}_1, \boldsymbol{W}_2 \}$ and $\boldsymbol{P} = \{ \boldsymbol{P}_1, \boldsymbol{P}_2 \}$. Following this approach, the complexity is reduced, as the gradients extract higher-order information for the various variables. 

\subsection{Gradient-based Meta Learning (GML) Architecture}
Conventional data-driven meta-learning methods typically demand extensive offline pre-training followed by further online adaptation and fine-tuning, as seen in model-agnostic meta-learning. This process often causes substantial fluctuations in data distribution and incurs high energy consumption due to large-scale training and adaptation requirements. Such limitations make these methods unsuitable for dynamic environments and latency-sensitive applications. To overcome these challenges, we propose a model-driven meta-learning framework that removes the need for pre-training while maintaining strong robustness. Instead of optimizing individual parameters, it concentrates on refining the optimization trajectory itself. In this regard, our meta-learning optimization framework can be defined by three iterative layers. These layers are formulated as follows:
\subsubsection{Inner Iteration} 
In this iteration, we optimize the variables $\boldsymbol{W}$ and $\boldsymbol{P}$ in a cyclic manner. This is done by assigning each optimization variable a sub-neural network:
\begin{itemize}
  \item \textbf{Beamforming Vectors Network (BVN)}: In this network, the transmit beamforming vectors at the two BSs are optimized.  
\item \textbf{Pinching Positions Network (PPN)}: This network optimizes the pinching locations for BS1 and BS2.  
  \end{itemize}
For the aforementioned sub-networks, a sequential update for the variables is adopted, in which it starts from preselected initial points. This process is defined as follows:
\begin{align}
&\boldsymbol{W}^*=R(\boldsymbol{W}^{(i,j)},\boldsymbol{P}^*), \\
   & \boldsymbol{P}^*=R(\boldsymbol{P}^{(i,j)},\boldsymbol{W}^*),
\end{align}
where the terms $\boldsymbol{W}^{(i,j)}$ and  $\boldsymbol{P}^{(i,j)}$ are values of the variables $\boldsymbol{W}$ and $\boldsymbol{P}$ at the inner iteration $i$ and outer iteration $j$ of the optimization algorithm. The details about the two sub-networks are provided below: 
\par \textit{\textbf{Beamforming Vectors Network (BVN)}}: This network is utilized to optimize the beamforming vectors at both BSs in the $i$-th inner iteration and $j$-th outer iteration with the aim of maximizing the objective function $R(\boldsymbol{W}^{(i,j)},\hat{\boldsymbol{P}})$ with $\hat{\boldsymbol{P}}$ being the updated or the initial value for the PAs locations. To accomplish this, the gradient of $\boldsymbol{W}^{(i,j)}$ is evaluated and fed to the neural network that will output $\Delta \boldsymbol{W}^{(i,j)}$ added to $\boldsymbol{W}^{(i,j)}$ in the following manner:
\begin{equation}
\begin{split}
&\boldsymbol{W}^{(i+1,j)}=\boldsymbol{W}^{(i,j)}+\Delta\boldsymbol{W}^{(i,j)}. \label{pup}
\end{split}
\end{equation}
\par \textit{\textbf{Pinching Positions Network (PPN)}}: Similar to the strategy of \textbf{BVN}, we define the objective function at the $i$-th inner iteration and $j$-th outer iteration as $R(\boldsymbol{P}^{(i,j)}, \hat{\boldsymbol{W}})$, where $\hat{\boldsymbol{W}}$ denotes the initial or updated beamforming matrix of the BSs. After computing the gradient of $\boldsymbol{P}$ with respect to the achievable sum rate, the neural network generates an update $\Delta \boldsymbol{P}^{(i,j)}$, which is then added to $\boldsymbol{P}^{(i,j)}$ as follows:
\begin{equation}
\boldsymbol{P}^{(i+1,j)} = \boldsymbol{P}^{(i,j)} + \Delta \boldsymbol{P}^{(i,j)}.
\label{cup}
\end{equation}

\begin{algorithm}[!t]
\footnotesize
\caption{Proposed algorithm}\label{alg:3}
 Randomly initialize $\theta_{\textbf{P}},\theta_{\textbf{W}},\textbf{P}^{(0,1)}$, $\textbf{W}^{(0,1)}$;\\
 Initialize inner iterations, $N_i$, outer iterations, $N_o$, and epoch iterations, $N_e$;\\
\For {$e={1,2,...,N_e}$}{
$\bar{\mathcal{L}}=0$;\\
MAX$=0$;\\
\For {$j=1,2,...N_o$}{
  $\textbf{W}^{(0,j)}=\textbf{W}^{(0,1)}$ ;\\
   $\textbf{P}^{(0,j)}=\textbf{P}^{(0,1)}$;\\

      \For {$i= 1,2,...,N_i$}{
      $R_{\textbf{W}}^{(i-1,j)}=R(\textbf{W}^{(i-1,j)},\textbf{P}^*)$;\\
      $\Delta \textbf{W}^{(i-1,j)}$= $\textbf{\textrm{BVN}}(\Delta_{\textbf{W}}R_{\textbf{W}}^{(i-1,j)})$;\\
      $\textbf{W}^{(i,j)}=\textbf{W}^{(i-1,j)}+\Delta \textbf{W}^{(i-1,j)}$;
      }
      $\textbf{W}^*=\textbf{W}^{(N_i,j)}$;\\
        \For {$i= 1,2,...,N_i$}{
      $R_{\textbf{P}}^{(i-1,j)}=R(\textbf{P}^{(i-1,j)},\textbf{W}^*)$;\\
      $\Delta \textbf{P}^{(i-1,j)}$= $\textbf{\textrm{PN}}$
      $(\Delta_{\textbf{P}}R_{\textbf{P}}^{(i-1,j)})$;\\
      $\textbf{P}^{(i,j)}=\textbf{P}^{(i-1,j)}+\Delta \textbf{P}^{(i-1,j)}$;
      }
      $\textbf{P}^*=\textbf{P}^{(N_i,j)}$;\\
      Apply the normalization as in \eqref{normalized}\\
   $\bar{\mathcal{L}}=\bar{\mathcal{L}}+\mathcal{L}^j$;\\
   \textbf{If} {$-\mathcal{L}^j>MAX$}\\
        \quad MAX=$-\mathcal{L}^j$;\\
        \quad $\textbf{W}_{opt}=\textbf{W}^*$;\\
        \quad $\textbf{P}_{opt}=\textbf{P}^*$;\\
   \textbf{end if}
   }
   
$\bar{\mathcal{L}}=\frac{1}{N_o}*\bar{\mathcal{L}}$;\\
update $\theta_{\textbf{P}}$ , $\theta_{\textbf{W}}$, \\

}
return $\textbf{W}_{opt},\textbf{P}_{opt},$
\end{algorithm}

At each outer iteration, a total of $N_o$ inner iterations are performed to accumulate the meta-loss. This meta-loss function is designed to jointly account for the objective of maximizing the achievable sum rate while ensuring that all feasibility constraints of the optimization problem \eqref{prob:P1} are satisfied. Accordingly, the meta-loss function is expressed as follows:
\begin{equation}
    \mathcal{L}^j= \mathcal{L}_{rate}^j + \mathcal{L}_{threshold}^j + \mathcal{L}_{PA1}^j +\mathcal{L}_{PA2}^j, 
    \label{loss}
\end{equation}
where the first term $\mathcal{L}_{rate}^j$ in the equation represents the loss corresponding to the objective function of \eqref{prob:P1} and is expressed as follows:
\begin{equation}
    \mathcal{L}_{rate}^j =  - \sum_{k \in \mathcal{K}} R_k.
\end{equation}
Now, in order to guarantee achieving the QoS requirements of the users, the term $\mathcal{L}_{threshold}^j$ is defined as:
\begin{equation}
    \mathcal{L}_{threshold}^j= \sum_{k \in \mathcal{K}} \zeta_1 \lambda_1 (R_k), 
\end{equation}
where a regularization term $\zeta_1$ is added and the indicator function $\lambda_1(.)$ is expressed as:
\begin{equation}
    \lambda  = \begin{cases}0, & \mbox{if } \mbox{$R_{th}-R_k \leq 0$}, \\  1, & \mbox{} \mbox{otherwise}. \end{cases}
\end{equation}
Next, to ensure the minimum antenna separation, we add the term $ \mathcal{L}_{PA1}^j$ that is defined as:
%N_{R_c}
\begin{equation}
%\footnotesize
    \mathcal{L}_{MA}^{j} = \sum_{n_{1} = 0 }^{N_{1}} \sum_{n_{2} = 0 }^{N_{2}}\zeta_2 \lambda_2 (\mathcal{A}_{n_1}+\mathcal{A}_{n_{2}}),
\end{equation}
\begin{align}
%\footnotesize
&\mathcal{A}_{n_1}= \nonumber \\
   &DS-||\textbf{p}_{n_1,p}^{BS}-\textbf{p}_{n_1,p'}^{BS}|| \le 0, 1 \le p \neq p' \le P_{n_1}, \forall n_1 \in \mathcal{N}_1,
   \label{ma1}
\end{align}
\begin{align}
%\footnotesize
&\mathcal{A}_{n_2}= \nonumber \\
   &DS-||\textbf{p}_{n_2,p}^{BS}-\textbf{p}_{n_2,p'}^{BS}|| \le 0, 1 \le p \neq p' \le P_{n_2},
   \forall n_2 \in \mathcal{N}_2,
   \label{ma1}
\end{align}
where the indicator function outputs 0 if the term :
\begin{equation}
\Omega \;\equiv\;
\max\!\left\{
\max_{\,n_1 \neq \hat{n}_1}\mathcal{A}_{n_1},\;
\max_{\,n_{2} \neq \hat{n}_{2}}\mathcal{A}_{n_{2}},
\right\} \;\le\; 0
\notag
\end{equation}
holds and yields 1 otherwise. $\zeta_2$ is a regularization term. Finally, to handle the last constraint \eqref{c3}, we add the term $\mathcal{L}_{PA2}^j$, where it is defined in a manner similar to the previous constraints. 
Although the terms introduced in $\eqref{loss}$ handle the constraints \eqref{c1} - \eqref{c3}, the power budget constraint at the BSs is still not tackled. In light of this, we leverage the normalization technique introduced in \cite{loli2024metalearningbasedoptimizationlarge}: 
\begin{equation} U(\boldsymbol{W}_b^*) = \begin{cases}\boldsymbol{W}_b^*, & \mbox{if } \mbox{$P_b^* \leq P_{BS,b}$}, \\ \sqrt{\frac{P_{BS,b}}{Tr(\boldsymbol{W}_b^*(\boldsymbol{W}_b^*)^H) }}\boldsymbol{W}^*_b, & \mbox{otherwise}, \end{cases} \label{normalized} \end{equation}
\noindent where $P_b^* = Tr(\boldsymbol{W}_b^*(\boldsymbol{W}_b^*)^H)$.

%represent the power corresponding to the beamforming matrix $\boldsymbol{W}_b^*$.
\subsubsection{Epoch Iteration} We update the neural network parameters in this level. It involves $N_e$ outer iterations, then accumulating and averaging the losses as follows:
\begin{equation}
    \bar{\mathcal{L}}=\frac{1}{N_e}\sum_{j=1}^{N_e}\mathcal{L}^j.
\end{equation}
Subsequently, backpropagation is applied, and the Adam optimizer is utilized to update the parameters of the neural networks within the proposed model, as given below:
\begin{equation}
\theta_{\textbf{W}}^* = \theta_{\textbf{W}} + \beta_{\textbf{W}} \cdot \textrm{Adam}(\Delta_{\theta_{\textbf{W}}}\bar{\mathcal{L}}, \theta_{\textbf{W}}), \label{w}
\end{equation}
\begin{equation}
\theta_{\textbf{P}}^* = \theta_{\textbf{P}} + \beta_{\textbf{P}} \cdot \textrm{Adam}(\Delta_{\theta_{\textbf{P}}}\bar{\mathcal{L}}, \theta_{\textbf{P}}), \label{p}
\end{equation}
where $\theta_{\textbf{W}}$, and $\theta_{\textbf{P}}$ denote the parameters of the sub-networks \textbf{BVN}, and \textbf{PPN}, respectively. Their corresponding learning rates are represented by $\beta_{\textbf{W}}$, and $\beta_{\textbf{P}}$. The overall GML architecture is depicted in Fig. \ref{GML_Model} while the training procedure is summarized in \textbf{Algorithm 1}.
\begin{table}[t]
\caption{Simulation Parameters}
\centering
\begin{tabular}{||l||c||l||c||}
    \hline
        Parameters & Values & P. & Values \\ \hline
        ~$P_{BS,b} ,\forall b \in \{1, 2 \}$ & ~18 dBm& ~$\lambda$ & ~0.01 \\ \hline
        ~$\mathcal{D}_{n_b} \forall b \in  \{1, 2\}$ & ~80m & ~$\sigma^2_k$ & -173 dBm/Hz \\ \hline
        ~$P_{n_b} \forall b \in \{1, 2 \}$  & ~4 & ~$A_{n_1}$ & ~10m \\ \hline
        ~$N_b \forall b \in \{1, 2 \}$ & ~2 & ~$A_{n_2}$ & ~15 \\ \hline
        ~$R_{th}$ & ~0.2 bps/Hz & ~$\alpha$ & ~3.9\\ \hline
        ~$K$ & 2 & ~$\beta_{\textbf{W}}$ & ~$1e^{-3}$\\ \hline
        ~$\beta_{\textbf{P}}$ & ~$1.6e^{-3}$ & ~$\zeta_2$ & ~$0.01$\\ \hline
        ~$\zeta_1$ & ~$0.0001$ & ~ & ~\\ \hline
  \end{tabular} 
  \label{T1}
\end{table}
\begin{table}
\centering
\caption{Number of neurons in the neural networks}
\begin{tabular}{ |l|c|c|c| }
 \hline
  \textbf{Layer name} & \textbf{BVN} & \textbf{PPN} \\
  \hline
Input Layer & 2K+2 & K \\
 \hline
Linear Layer  & 210 & 210  \\
 \hline
 ReLU Layer & 210 & 210 \\
 \hline
Output Layer & 2K+2 & K \\
 \hline
\end{tabular}
\end{table}
\begin{figure*}
  \centering
  \subfigure[Convergence of the solution approach.]{\includegraphics[width=0.30\linewidth]{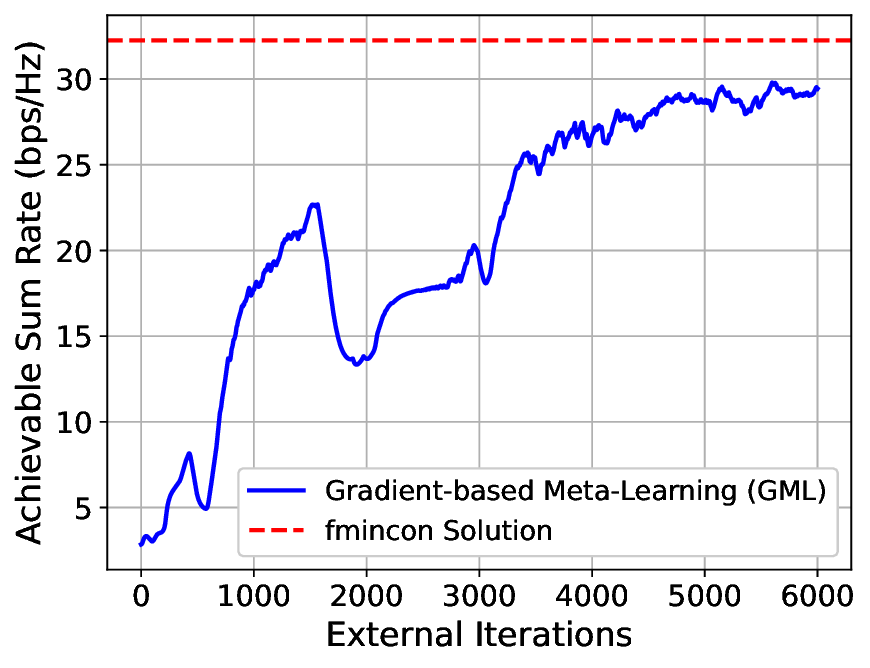}}
  \subfigure[BS power vs achievable sum rate.]  {\includegraphics[width=0.33\linewidth]{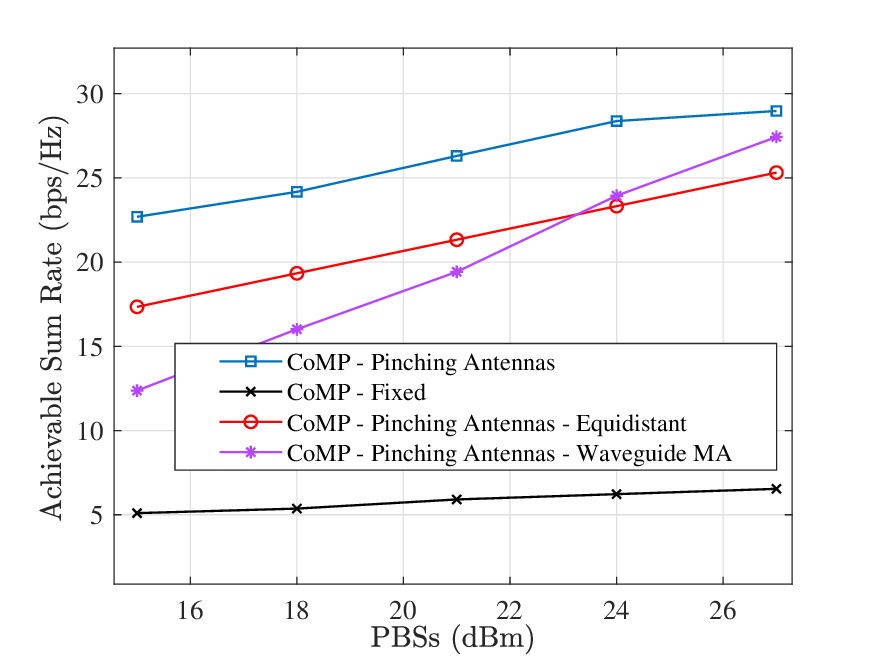}}
  \subfigure[Rate threshold vs achievable sum rate.]{\includegraphics[width=0.33\linewidth]{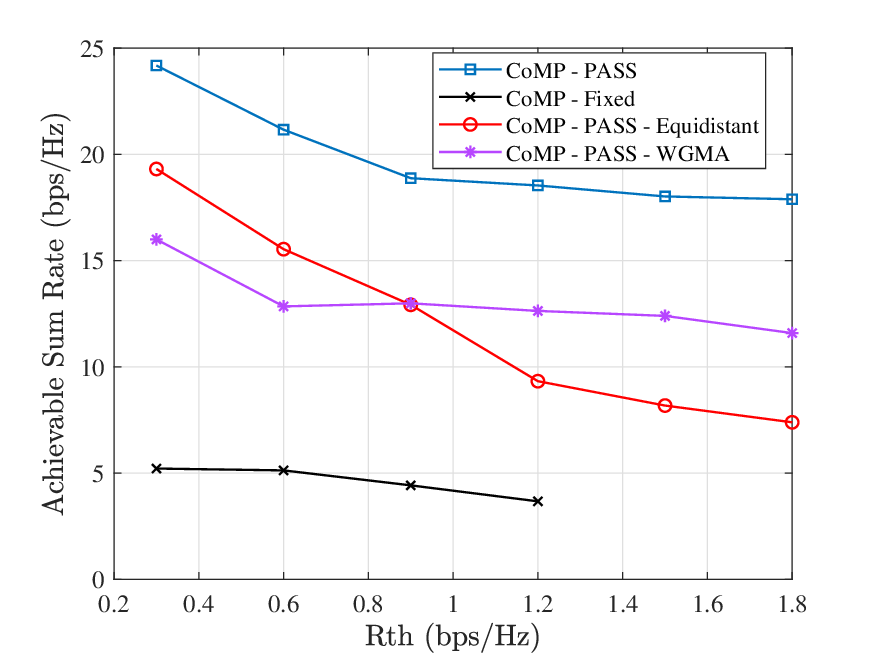}}
  \setlength{\belowcaptionskip}{-12pt} 
  \caption{Numerical results for the model under different parameters.}
  \label{three_figures}
  \vspace{-0.3cm}
\end{figure*}
\section{Numerical analysis}
The numerical evaluation of the CoMP-PA model is presented in this section. The performance is compared with respect to a set of benchmarks defined as follows:
\begin{itemize}
  \item \textbf{CoMP - Fixed}: This model comprises two BSs equipped with traditional uniform linear arrays (ULA).
\item \textbf{CoMP - Pinching Antennas - Equidistant}: This benchmark follows a similar architecture to the presented model, with PAs being positioned in an equidistant manner along the waveguides.  
\item \textbf{CoMP - Pinching Antennas - Waveguide division multiple access (WDMA)\cite{zhao2025waveguidedivisionmultipleaccess}}: In this benchmark, we follow the waveguide division multiple access technique in which each user is assigned a waveguide from BS1 and another one from BS2. However, the transmission of other waveguides is considered as interference. 
\end{itemize}
Unless stated otherwise, the main simulation parameters are summarized in Table I, while the configurations of the neural networks are presented in Table II. We examine 2 users that are uniformly distributed along the spanning area of the 2 BSs (80m $\times$ 80m). 

In Fig. \ref{three_figures} (a), we analyze the convergence behavior of the proposed meta-learning algorithm. The figure illustrates the achievable sum rate of the users as a function of the outer iterations within a single epoch. As the number of iterations increases, the achievable rate exhibits a steady improvement until around 1500 iterations, where a temporary drop occurs. This decline can be attributed to the model’s effort to maximize the objective function while satisfying the imposed constraints, which introduces penalties when violations occur. Subsequently, the sum rate continues to rise until convergence is achieved at approximately 30 bps/Hz, confirming the effectiveness of the proposed model by achieving about 92.6\% compared to the non-convex optimization solver fmincon that can provide a very close to optimal result as presented in the red line \cite{9586734}. 

Fig. \ref{three_figures} (b) examines the achievable sum rate versus the power budget at the two BSs. It is clear that with the increase in the budget, the rate increases for the different benchmarks, with superiority for the presented model due to the ability of PAs to be flexibly positioned, creating a clear LoS with the communication users. Also, the presented model showed a higher upper bound compared to the traditional CoMP system that utilizes ULAs, proving a better ability in managing intra-user interference. Moreover, the advantage of the presented model over the same system that relies on equidistant placement for the pinching locations proves the importance of optimizing PA placement to control the channel between the waveguides and the users. In addition, the CoMP model incorporating PAs with WDMA demonstrates a performance comparable to the proposed model at 27 dBm, while surpassing the model with equidistant PA placement. This improvement can be attributed to its enhanced capability to mitigate interference from the waveguides serving other users as the transmit power of the BSs increases.

Fig. 3 (c) shows the achievable sum rate versus the QoS requirements of the users. The graph clearly demonstrates that the increase in the threshold leads to a decrease in the user rate for the different benchmarks. This can be explained by the need of the BSs to allocate more power to some users to meet their QoS requirements, losing the advantage of maximizing the sum rate for others. Moreover, the suggested CoMP-PA model demonstrated a better performance compared to other benchmarks, especially the CoMP - Fixed (with ULAs), which started experiencing infeasibility after 1.2 bps/Hz. This is due to the ability of PAs to dynamically adjust their positions along the dielectric waveguides, enabling more effective alignment with user locations, minimizing path loss, and mitigating inter-user interference. Consequently, the system achieves enhanced spatial multiplexing gains and improved overall spectral efficiency, even under high-rate transmission.
\section{Conclusion}
In this work, we investigated a CoMP system employing PASS to serve communication users using the SDMA technique. We formulated an optimization problem that aims to maximize the achievable sum rate through the joint design of transmit beamforming vectors and pinching locations, while ensuring users’ QoS constraints. To tackle the challenge of the non-convexity in the generated problem, a GML framework was developed, offering a scalable and efficient solution. The numerical results verified the effectiveness of the proposed method, achieving up to 92\% of the optimal performance and demonstrating clear superiority over existing benchmark schemes, adding to the ability to overcome the upper-bound limitations in conventional CoMP systems.

%\vfill
\bibliographystyle{IEEEtran}
\bibliography{ref}
\end{document}